\def\ARXIV{1}
\documentclass[11pt]{article}

\usepackage[letterpaper, margin=1in]{geometry}
\usepackage{amsmath, amssymb}
\usepackage{booktabs}
\usepackage{multirow}
\usepackage{graphicx}
\usepackage{xcolor}
\usepackage[hidelinks]{hyperref}
\usepackage{microtype}
\usepackage{authblk}
\usepackage{caption}
\usepackage{titlesec}
\usepackage{enumitem}
\usepackage{url}
\usepackage[most]{tcolorbox}

\titleformat*{\section}{\large\bfseries}
\titleformat*{\subsection}{\normalsize\bfseries}

\newcommand{\tierC}[0]{\textbf{\textcolor[HTML]{1a7a3a}{CONFIRMED}}}
\newcommand{\tierE}[0]{\textbf{\textcolor[HTML]{8a6d00}{EXPLORATORY}}}
\newcommand{\tierR}[0]{\textbf{\textcolor[HTML]{a32424}{REFUTED}}}
\newcommand{\tierI}[0]{\textbf{\textcolor[HTML]{555555}{INCONCLUSIVE}}}
\newcommand{\tierD}[0]{\textbf{\textcolor[HTML]{2a4d9c}{DEMONSTRATED}}}

\title{\textbf{The Judge Knows When It Knows:\\
Calibrated Abstention for LLM-Based A/B-Test Prediction}}

\author[1]{Tyler Dooskin}
\author[1]{the Squoosh Technical Staff}
\affil[1]{Squoosh \quad \texttt{tyler@squoosh.ai}}
\affil[ ]{\normalfont\small This report reflects the work of Squoosh's engineering and research team; individual contributors are credited collectively at their preference.}

\date{June 2026}

\begin{document}
\maketitle

\begin{abstract}
\noindent
Can a multimodal LLM predict which version of a web page will win a real A/B test from screenshots alone? We report the most complete answer we are aware of, from six weeks of pre-registered experiments on real conversion tests: \textbf{mostly no --- and the exceptions are precisely identifiable in advance.}

Three findings structure the paper. \textbf{(1) Unconditional winner prediction does not clear the honesty bar.} On 330 real A/B tests, a Gemini~3 Flash judge attains Cohen's $\kappa = 0.141$ $[0.034, 0.248]$ --- detectably above chance, but on the trustworthy (statistically significant) half of the labels the evidence is inconclusive ($\kappa = 0.108$ $[-0.049, 0.264]$). We further show that \textbf{44\% of the ``ground-truth'' labels in our primary corpus --- the curated case library of a leading CRO agency --- come from non-significant tests}, and that the judge agrees \emph{more} with those unreliable labels than with reliable ones --- the signature of a shared prior between label curator and model, not of prediction. Every standard improvement lever fails to move this: a $2.8\times$ more expensive frontier model is statistically \emph{equivalent} to Flash as the judge ($\Delta\kappa = -0.037$ $[-0.152, +0.076]$, within a $\pm 0.20$ TOST bound, $n{=}159$ paired); prompt-mechanism redesign, stimulus fidelity, and change-type priors all fail their pre-registered gates. \textbf{(2) The judge's \emph{confident} calls are different.} Gating predictions on internal panel agreement concentrates real signal: at a vote-margin $\geq 0.6$ gate the judge calls 49\% of tests and abstains on the rest, and on significant-only labels the called subset reaches $\kappa = 0.311$ $[0.025, 0.555]$ --- the only operating point that survives our own label-validity audit (exploratory tier; a pre-registered held-out confirmation at the stricter unanimity point returned $\kappa = -0.100$ $[-0.200, 0.000]$ on an underpowered 33-call cohort, which by our locked decision rule upgrades nothing and kills nothing; a powered confirmation on partner data is pre-registered and in motion). The abstention behavior itself is robust: coverage replicates out-of-sample (27.8\% $\rightarrow$ 30.3\%), and the gated calls are perfectly reproducible across independent runs (31/31) --- which we show is exactly why reproducibility must never be mistaken for validity. The mechanism is measurable: judges differing in model or prompt agree with each other at $\kappa = 0.74$--$0.88$ while each agrees with outcomes at only $\sim 0.2$, and a 16-vote panel carries $\approx 2$ effective independent votes. \textbf{(3) The economics work.} Vote-subsampling shows a 9-vote panel preserves the gated operating point ($\kappa = 0.264$--$0.271$ at 46\% coverage) at roughly half the inference cost of the full panel, and the proxy tier runs at $\sim\$0.02$ per evaluation versus $\sim\$1.25$ for a browser-agent tier.

The deliverable this evidence supports is not a win-rate oracle. It is a \textbf{calibrated screening instrument}: confident-call-or-abstain winner reads, a grounded mobile-UX audit whose findings pass an adversarial verification pass (12 of 13 verified, 1 auto-rejected), and a published claims ledger in which \emph{every} number carries an evidence tier. For calibration of expectations: in the best published human benchmark, 200+ CRO professionals averaged 29\% on 8 real tests --- indistinguishable from chance; and a fresh \textbf{15-expert baseline on our own instrument reproduces the shared-prior collapse directly} --- experts agree with \emph{each other} (inter-rater $\kappa = 0.53$) but score at chance against the real \emph{outcomes} ($\kappa \approx 0$), with $n_{\mathrm{eff}} \approx 1.8$ effective independent experts of 15. We release our pre-registrations, locked gates, negative results, statistical harness, the human responses, and the full experiment ledger.
\end{abstract}

\ifdefined\ARXIV
\begin{tcolorbox}[colback=blue!4, colframe=blue!40, title=\textbf{Practical implications for experimentation teams}, fonttitle=\bfseries]
\small
\begin{itemize}[leftmargin=*, topsep=2pt, itemsep=2pt]
    \item Raw accuracy quoted against curated A/B catalogs is a base-rate artifact: the rule \emph{always pick the variant} scores $\sim$75--85\% on such catalogs while contributing zero decision value. Evaluations of this task should report chance-corrected metrics ($\kappa$), stratified by \emph{statistically significant} labels.
    \item Expert humans are not the bar one might assume: in the largest public test of practitioners (200+ professionals, 8 real A/B tests, prize money on the line), the average was chance. A judge that knows \emph{when} it knows is the relevant comparison point for screening layers.
    \item What this evidence supports deploying: \textbf{(a)} confident-call-or-abstain winner reads --- the judge panel calls a winner only when it clears its confidence gate, and abstains otherwise; \textbf{(b)} screenshot-grounded UX audits whose findings pass an adversarial verification pass before being surfaced; \textbf{(c)} a tiered claims ledger in which every externally quoted number carries its evidence tier, including failed experiments.
    \item What it does not support claiming: predicted conversion-lift percentages, ``mobile-validated'' accuracy (no mobile-rendered labeled corpus exists, ours or anyone's), or any unconditional accuracy claim. The ledger in Appendix~A makes this binding.
\end{itemize}
\end{tcolorbox}
\else
\begin{tcolorbox}[colback=blue!4, colframe=blue!40, title=\textbf{Why this matters if you run experiments for a living}, fonttitle=\bfseries]
\small
\begin{itemize}[leftmargin=*, topsep=2pt, itemsep=2pt]
    \item Any vendor can quote ``$\sim$80\% accuracy'' on curated A/B catalogs --- so can the rule \emph{always pick the variant}, because curated catalogs are $\sim$75--85\% variant-wins. Raw accuracy on this task is a base-rate artifact. Demand chance-corrected metrics ($\kappa$), and demand them on \emph{statistically significant} labels.
    \item Expert humans are not the bar you think: in the largest public test of practitioners (200+ professionals, 8 real A/B tests, prize money on the line), the average was chance. A judge that knows \emph{when} it knows beats both naive AI and unaided intuition as a screening layer.
    \item What Squoosh ships on this evidence: \textbf{(a)} a confident-call-or-abstain winner read --- when our judge panel clears its confidence gate it calls a winner, otherwise it tells you it doesn't know; \textbf{(b)} a mobile-UX audit of concrete, screenshot-grounded problems, each adversarially re-verified before you see it; \textbf{(c)} this paper's claims ledger --- our numbers, with their tiers, including the experiments that failed.
    \item What we will not sell you: predicted conversion-lift percentages, ``mobile-validated'' accuracy (no mobile-rendered labeled corpus exists --- ours or anyone's), or any unconditional accuracy claim. The ledger in Appendix~A is the contract.
\end{itemize}
\end{tcolorbox}
\fi

\section{Introduction}

LLM judges are being sold as pre-test oracles for conversion optimization: show the model two page designs, get tomorrow's A/B winner today. The pitch is seductive because the evaluation is easy to game --- often unintentionally. Curated A/B-test catalogs are heavily skewed toward variant wins (teams build variants they expect to win, and curators showcase wins), so a judge that simply prefers the changed page scores $\sim$75--85\% ``accuracy'' while contributing zero decision value. The failure mode of this product category is not weak models; it is \textbf{unfalsifiable marketing built on base-rate exploitation}.

This paper is our attempt to hold ourselves to a falsifiable standard, in public. Over six weeks we ran a pre-registered experimental program against real A/B outcomes --- locked hypotheses, locked decision gates, locked prediction files, adversarial verification of our own headline results --- and we report everything: the negative results (most of the program), the one operating point that survives, its held-out test that came back inconclusive, and the powered confirmation now in motion on partner data. Three properties distinguish this from a standard vendor whitepaper:

\begin{enumerate}[leftmargin=*, itemsep=2pt]
    \item \textbf{Every number carries an evidence tier} (\S\ref{sec:tiers}; ledger in Appendix~A). ``Confirmed'' means pre-registered, powered, and gate-passed. Nothing exploratory is dressed up as confirmed --- including our own headline.
    \item \textbf{The machinery caught our own best results.} Our label-validity audit partially invalidated our strongest unconditional benchmark; our held-out confirmation declined to promote our strongest gated estimate. Both events are reported here as findings, because a measurement system that cannot kill its owner's claims is advertising, not measurement.
    \item \textbf{The product claim is calibrated, not maximal.} The instrument we ship abstains by design on $\sim$half of inputs. We argue --- with a frontier of operating points, not a single cherry-picked threshold --- that a small number of high-precision calls plus honest abstention is worth more to an experimentation program than a confident answer to every question.
\end{enumerate}

\subsection{The claims discipline}\label{sec:tiers}

We tag every quantitative claim with one of six tiers, defined once and used everywhere (including in our sales materials --- the tags are the contract):

\begin{itemize}[leftmargin=*, topsep=2pt, itemsep=1pt]
    \item \tierC{} --- pre-registered hypothesis, adequately powered, locked gate passed.
    \item \textbf{REPLICATED-DIRECTIONAL} --- same direction in $\geq 2$ independent runs; no single powered held-out confirmation yet.
    \item \tierE{} --- post-hoc, in-sample, or threshold-selected; selection effects not fully modeled.
    \item \tierI{} --- a pre-registered held-out test that returned an underpowered null: upgrades nothing, refutes nothing.
    \item \tierR{} --- tested and failed: CI includes zero on trustworthy data, or contradicted by a stronger design.
    \item \tierD{} --- engineered pipeline behavior, verified end-to-end; not a statistical-skill claim, carries no CI.
\end{itemize}

\section{Related work}\label{sec:related}

\textbf{LLM-as-judge and its failure modes.} Using LLMs to evaluate artifacts at scale was popularized by MT-Bench and Chatbot Arena \cite{zheng2023judging} and framing-based evaluators such as G-Eval \cite{liu2023geval}. The judge paradigm's known pathologies --- position bias, verbosity bias, self-preference --- are documented by Wang et al.\ \cite{wang2023fair} among others; our orientation-swap control (\S\ref{sec:stack}) is the standard mitigation, and our panel design follows the multi-judge direction of Verga et al.\ \cite{verga2024poll}, though our results add a caution that panel \emph{agreement} can be deterministic prior-expression rather than independent evidence (\S\ref{sec:heldout}). Our task differs from typical judge applications in one decisive way: the ground truth is an \emph{external causal outcome} (a randomized experiment's winner), not a human preference label, so judge--label agreement cannot be improved by aligning the judge to the labeler's taste --- and shared priors between judge and labeler become a confound rather than a convenience (\S\ref{sec:labels}).

\textbf{Selective prediction and calibration.} Prediction with a reject option dates to Chow \cite{chow1970}, with the modern risk--coverage formulation developed by El-Yaniv and Wiener \cite{elyaniv2010} and brought to deep networks by Geifman and El-Yaniv \cite{geifman2017}. Kamath et al.\ \cite{kamath2020} apply selective prediction to QA under domain shift; Guo et al.\ \cite{guo2017} establish that modern networks are systematically miscalibrated. Our contribution to this line is empirical rather than algorithmic: in a domain where unconditional skill is statistically unprovable on trustworthy labels, a vote-margin gate over a counterbalanced panel isolates the subset of inputs where measurable skill exists --- and a pre-registered held-out test of that gate illustrates how easily selective-prediction gains can fail to transfer (\S\ref{sec:heldout}).

\textbf{Online controlled experiments.} The methodology and ground truth of our corpus come from the A/B-testing literature: Kohavi et al.\ \cite{kohavi2009,kohavi2020} codify trustworthy experimentation practice, and Kohavi and Thomke \cite{kohavi2017hbr} report that even at mature experimentation organizations only 10--20\% of expert-designed treatments win --- the base rate that makes screenshot-level winner prediction information-poor. The practitioner-prediction benchmark we use as a human anchor (200+ professionals, 8 real tests, mean 2.3/8) is reported in Kohavi's practitioner materials accompanying \cite{kohavi2020}.

\textbf{LLM user simulation.} A complementary line uses LLM agents to \emph{simulate users} rather than judge outcomes: generative agents \cite{park2023generative} and, closest to our setting, UXAgent's LLM-agent usability testing for web design \cite{lu2025uxagent}. Our production system runs synthetic shoppers in real browsers in exactly this spirit; this paper isolates the judging question --- what the simulation layer can honestly \emph{claim} about real experimental outcomes --- and supplies the calibration discipline that any user-simulation product ultimately needs.

\textbf{Statistical instrumentation.} Cohen's $\kappa$ \cite{cohen1960}, percentile bootstrap \cite{efron1993}, TOST equivalence testing \cite{schuirmann1987}, and pre-registration \cite{nosek2018} are each standard; our assembly of them --- locked gates committed before data, equivalence bounds declared in advance, gate-on-resample intervals for selective metrics --- is, to our knowledge, not yet standard in commercial LLM-evaluation practice, and Appendix~\ref{app:protocol} specifies it precisely so it can be.

\section{The Reliability Stack}\label{sec:stack}

All results below run through one statistical harness, reused unchanged across experiments:

\begin{itemize}[leftmargin=*, topsep=2pt, itemsep=1pt]
    \item \textbf{Chance-corrected gating.} Cohen's $\kappa$ \cite{cohen1960} is the gate metric everywhere; raw accuracy is reported but never gates (\S\ref{sec:baserate} shows why with a worked example where the judge's most-confident slice scores 75.8\% accuracy and still has \emph{negative} $\kappa$).
    \item \textbf{Percentile bootstrap CIs} \cite{efron1993}, $B = 10{,}000$, fixed seed $42$; \emph{paired joint} resampling for any two-arm contrast on shared test IDs (both arms resampled together per replicate).
    \item \textbf{Gate-on-resample for gated metrics.} When a confidence gate selects the scored subset, we resample the \emph{full} test set and re-apply the gate inside every bootstrap replicate, propagating coverage uncertainty into the $\kappa$ CI. Conditioning on the realized called subset --- the natural shortcut --- understates the interval.
    \item \textbf{Pre-registration with locked gates \cite{nosek2018}.} Hypotheses, decision rules, power analyses, and verdict colors (GREEN/YELLOW/RED/INCONCLUSIVE) are committed to the repository \emph{before} predictions are collected; prediction files and analysis outputs are locked after.
    \item \textbf{Position counterbalancing.} Judge panels see both image orders; an orientation-swap control exposed a content-independent second-image position bias early in the program.
    \item \textbf{Trivial baselines as floors.} Always-predict-B, pixel-difference counters, and DOM-feature heuristics run alongside every benchmark.
    \item \textbf{Adversarial verification of our own results.} Headline findings get an explicit falsification pass (independent re-derivation from raw votes, per-test diffs against file collisions, label-validity stratification) before they are reported --- this pass is what produced \S\ref{sec:labels} and \S\ref{sec:heldout}.
\end{itemize}

\section{Data: real A/B tests, and the label-validity census}\label{sec:labels}

\subsection{Corpus}

Our primary corpus is the proprietary case library of a \textbf{leading conversion-rate-optimization (CRO) agency}, provided under a data-use agreement. The provider requested anonymity --- a standard arrangement for commercial datasets --- so we follow the usual convention: we describe the source rather than name it, state the terms under which it was obtained, and report every corpus property a reviewer needs to assess validity (see also the data-availability terms in \S\ref{sec:repro}). The library catalogs real A/B tests run on live commercial sites: 546 unique tests after canonicalization, 337 with a decisive recorded winner (80 control/A wins, 257 variant/B wins --- a 76\% B-prior), with per-arm sample sizes and successes for 520, enabling exact recomputation of each test's significance. Screenshots for both arms of all tests are archived with content-hash provenance. The powered benchmark set is the 330--332 decisive tests with renderable screenshots across desktop-only, mobile-only, and blended segments. The judge is Gemini~3 Flash, multimodal, with a counterbalanced 16--20-vote panel per test; the production prompt (\texttt{v4}) survived four generations of pre-registered prompt-revision gates.

\subsection{The census: 44\% of ``ground truth'' is editorial}

Recomputing significance from the published per-arm counts on the 336 decisive labels used by our powered benchmark:

\begin{center}
\begin{tabular}{lrr}
\toprule
Label class & $n$ & A-wins \\
\midrule
Significant ($p < .05$, recomputed) & 152 & 30 \\
\textbf{Non-significant} (median recomputed $p = 0.13$) & \textbf{148} & 45 \\
No per-arm statistics published & 36 & 5 \\
\bottomrule
\end{tabular}
\end{center}

Nearly half the catalog's winner labels are directional point-estimate calls on tests that did not reach significance. Stratifying the judge's performance by label quality:

\begin{center}
\begin{tabular}{lcc}
\toprule
Benchmark set & all-labels $\kappa$ & significant-only $\kappa$ \\
\midrule
Powered pooled ($n{=}330/150$) & $0.141\ [0.034, 0.248]$ & $\mathbf{0.108\ [-0.049, 0.264]}$ \\
Replication arm ($n{=}159/78$) & $0.214\ [0.054, 0.365]$ & $\mathbf{0.120\ [-0.096, 0.340]}$ \\
\bottomrule
\end{tabular}
\end{center}

\textbf{On trustworthy labels, the evidence for unconditional judge skill is statistically inconclusive.} The direction is the diagnostic part: if non-significant labels were merely noisy, $\kappa$ against them would be \emph{lower}, not higher. A judge that agrees \emph{more} with unreliable labels than reliable ones is exhibiting a \textbf{shared-prior artifact}: on marginal tests the provider's directional call plausibly encodes CRO folk wisdom about what \emph{should} win, and the LLM was trained on the same folk wisdom. Agreement there has a common cause; it is not prediction. (Tier: \tierE{} as a stratification, but it \emph{removes} the proof of skill from the all-labels number rather than adding a claim --- the asymmetry matters.)

\textbf{Implication beyond this paper:} any LLM-judge accuracy number computed against a curated A/B catalog --- ours or any vendor's --- inherits this artifact unless it stratifies by label significance. We recommend the stratified report become standard practice.

\subsection{Why $\kappa$, with a worked example}\label{sec:baserate}

On the held-out cohort of \S\ref{sec:heldout}, the judge's most-confident (unanimous) slice is 93.9\% B-truth. A do-nothing \emph{always-B} rule scores 93.9\% accuracy there; the judge scores 75.8\% --- and its $\kappa$ is $-0.100$, because it called A six times (all wrong) and missed both real A-winners. High accuracy, negative skill, on the slice the judge was most confident about. Raw accuracy cannot see this; $\kappa$ is built to. Every gate in this program is therefore a $\kappa$ gate, and no accuracy figure in this paper is independently significant evidence of anything.

\section{The closed levers: powered negative results}\label{sec:negatives}

Each ``obvious fix'' below was tested under a locked pre-registration and failed its gate. We consider these results load-bearing: they are why the calibrated-abstention operating point (\S\ref{sec:abstention}) is the product, rather than a bigger model or a cleverer prompt.

\subsection{Model capacity: a frontier model is \emph{equivalent}, not better (\tierC{})}

Head-to-head as the judge on the same 159 tests (paired joint bootstrap): Gemini~3 Flash $\kappa = 0.214\ [0.054, 0.365]$ vs.\ Gemini~3.1~Pro $\kappa = 0.178\ [0.024, 0.327]$;
\[
\Delta\kappa(\text{Pro} - \text{Flash}) = -0.037\ [-0.152, +0.076].
\]
The CI includes zero \emph{and} lies entirely inside the pre-declared $\pm 0.20$ TOST-style equivalence bound \cite{schuirmann1987}: a powered \textbf{equivalence} claim, not an absence-of-evidence shrug. Pro also over-calls A more (calibration gap 11.3pp vs.\ 8.8pp) at $\sim$2.8$\times$ the latency and cost. Model capacity is not the bottleneck on this task.

\subsection{Prompt mechanism, stimulus fidelity, and priors (\tierR{})}

\begin{itemize}[leftmargin=*, topsep=2pt, itemsep=2pt]
    \item \textbf{Prompt redesign:} a calibration-targeted mechanism rewrite (M3) shrank the A-over-prediction gap (8.8\% $\rightarrow$ 7.0\%) but failed $\kappa$ non-inferiority against the incumbent ($0.173$ vs.\ $0.214$; paired $\Delta = -0.029\ [-0.10, +0.04]$). Four earlier prompt generations and two GEPA-style automated optimization chains either failed their holdout gates or won their gate while leaving corpus-level $\kappa$ at zero (full ledger in Appendix~B, including a cautionary finding: a reflection-LM data leak that caused candidate prompts to memorize benchmark pattern IDs as lookup tables, caught by a forensic audit and patched with a validator).
    \item \textbf{Stimulus fidelity:} judging real captured screenshots vs.\ reconstructed pages produced parity at the $\kappa \approx 0$ ceiling on matched tests ($\Delta\kappa = -0.07$, CI includes 0) --- fidelity was not the constraint.
    \item \textbf{Change-type priors:} leave-one-test-out pattern-level priors (``this category of change usually wins'') predict held-out outcomes at $\kappa = 0.015\ [-0.098, 0.131]$, versus the visual judge's $0.140\ [0.021, 0.259]$ on the same 279 tests. Within-pattern outcome consistency is poor; context dominates change type. The folk theory underlying most ``best-practices'' tools fails its own catalog.
    \item \textbf{Mobile:} on mobile-segment tests all judge arms sit at $\kappa \approx 0.00$--$0.02$ (CIs $\approx [-0.20, 0.27]$), and \emph{no mobile-rendered, outcome-labeled corpus exists anywhere we could find} to do better. We therefore make no mobile-accuracy claim, and flag any vendor who does. (The mobile capture-and-audit \emph{pipeline} is production-validated; the accuracy claim is what's unavailable.)
\end{itemize}

\section{The surviving signal: calibrated abstention}\label{sec:abstention}

\subsection{The frontier (\tierE{})}

The judge's panel votes define a natural confidence signal: the margin between votes for each arm. Gating on it trades coverage for precision:

\begin{center}
\begin{tabular}{lcccc}
\toprule
Gate & Coverage & $\kappa$ (all labels) & $\kappa$ (significant-only) \\
\midrule
none & 100\% & $0.141\ [0.034, 0.245]$ & $0.108\ [-0.049, 0.264]$ \\
margin $\geq 0.6$ & 49\% & $0.294\ [0.112, 0.470]$ & $\mathbf{0.311\ [0.025, 0.555]}$ \\
margin $\geq 0.8$ & 38\% & $0.301\ [0.080, 0.504]$ & $0.267\ [-0.045, 0.556]$ \\
unanimity & 25\% & $0.268\ [0.011, 0.515]$ & $0.165\ [-0.151, 0.492]$ \\
\bottomrule
\end{tabular}
\end{center}

The margin-$\geq 0.6$ row is the finding: \textbf{the judge's confident calls carry real signal even on trustworthy labels} --- the only cell in this research program where a significant-only CI excludes zero. The judge's \emph{forced} calls (no gate) do not. This independently re-derives, from a label-quality angle, the product decision to ship confident-call-or-abstain rather than always-answer. Honesty notes: the threshold is one of four examined (selection effect possible); the in-sample unanimity estimates on the calibration arms were higher ($\kappa = 0.349\ [0.044, 0.618]$ pooled at 27.8\% coverage) --- which is exactly why we ran a held-out confirmation.

\subsection{The held-out test, reported in full (\tierI{})}\label{sec:heldout}

We pre-registered a confirmation on the 109 decisive tests never scored by any calibration run, with a locked decision rule (unanimity gate only, no threshold sweep), the gate-on-resample bootstrap, and a locked verdict space that \emph{could not} return ``confirmed'' at this cohort size (33 expected unanimous calls vs.\ a $\sim$140-call power floor). Result:

\begin{center}
\small
\begin{tabular}{p{5.4cm}p{8.6cm}}
\toprule
Coverage (called / abstained) & 30.3\% (33 / 76) --- in the predicted 25--35\% band \\
$\kappa$ at unanimity & $\mathbf{-0.100\ [-0.200, 0.000]}$ \\
Called-subset acc.\ vs.\ always-B & 75.8\% vs.\ 93.9\% (see \S\ref{sec:baserate}) \\
Verdict per locked gate & INCONCLUSIVE (underpowered null: upgrades nothing, refutes nothing) \\
\bottomrule
\end{tabular}
\end{center}

The in-sample $\kappa \approx 0.35$ did not directionally replicate on this cohort; the cohort is also far too small to refute it (power $\approx 0.36$--$0.59$ against a true $\kappa = 0.30$). Per the locked rule we report it as exactly that. Two genuinely new findings came out of the error anatomy:

\begin{itemize}[leftmargin=*, topsep=2pt, itemsep=2pt]
    \item \textbf{Reliability is not validity.} The held-out run's unanimous calls agreed with an independent earlier run's calls on \textbf{31/31} overlapping tests (different panel size, different context framing) --- perfect test--retest reliability --- while scoring $\kappa = -0.100$ against truth. Unanimity in an LLM panel is \emph{one deterministic prior expressed many times}, not independent evidence accumulating. We offer this as a general caution for LLM-judge ensembles: agreement metrics cannot certify skill.
    \item \textbf{Confident errors cluster by folk wisdom.} The six wrong unanimous calls collapse onto four change-pattern families, and the flagship miss backed an interactive ``gradual engagement'' flow (36/36 votes across two panels) over the plain signup that actually won, significantly. The gate's failure mode is exactly the shared-prior artifact of \S\ref{sec:labels}, observed item-by-item.
\end{itemize}

\textbf{What would settle it:} $\sim$140--180 unanimous calls on real-outcome data the judge has never seen, from a corpus that does not carry the curated B-prior. A pilot protocol on a commerce A/B-testing platform partner's $\sim$300-test historical dataset is pre-registered and locked as of June~10; its verdict space and power ceiling are stated in the pre-registration \emph{before} data receipt.

\subsection{The prior crosses models and prompts, measured (\tierE{})}\label{sec:crossjudge}

If judge--label agreement on marginal tests reflects a shared training prior rather than prediction (\S\ref{sec:labels}), a sharp test is available: \emph{independently configured judges should agree with each other far more than either agrees with reality.} We have three complete 16-vote panels on the same blended corpus --- Flash+v4, Pro~3.1+v4 (different \emph{model}), and Flash+M3 (different \emph{prompt}) --- so the test costs nothing:

\begin{center}
\small
\begin{tabular}{lccc}
\toprule
Pair ($n$ shared tests) & inter-judge $\kappa$ & each vs.\ truth $\kappa$ & jointly-unanimous direction \\
\midrule
Flash vs.\ Pro (159) & $\mathbf{0.736\ [0.623, 0.841]}$ & $0.214$ / $0.178$ & 34/34 identical \\
v4 vs.\ M3 prompt (158) & $\mathbf{0.878\ [0.793, 0.947]}$ & $0.212$ / $0.183$ & 36/36 identical \\
\bottomrule
\end{tabular}
\end{center}

Two judges that differ in model or in prompt agree with \emph{each other} three to four times more strongly than either agrees with real outcomes. The shared-prior account stops being an inference here and becomes a measurement. Three corollaries:

\begin{itemize}[leftmargin=*, topsep=2pt, itemsep=2pt]
    \item \textbf{A panel vote is not a sample of independent evidence.} One-way ICC on the binary votes grouped by test gives $\rho = 0.34$--$0.51$ across the three judges: a 16-vote panel carries $n_{\mathrm{eff}} \approx 1.9$--$2.6$ \emph{effective independent votes}. This single number jointly explains why unconditional $\kappa$ is flat in panel size, why gated calls reproduce 31/31 across runs (\S\ref{sec:heldout}), and why unanimity must not be read as accumulating evidence.
    \item \textbf{Diversity does not rescue the task.} Mixed 8+8 panels (model-diverse or prompt-diverse, votes drawn hypergeometrically from the recorded tallies) perform within noise of pure panels at the unanimity gate (e.g., model-diverse $\kappa = 0.337$ $[0.235, 0.432]$ at 26\% coverage vs.\ pure-Flash $0.310$ at 25\%). Juries of correlated jurors are still one juror \cite{verga2024poll}.
    \item \textbf{A modest, honest ensemble gain exists.} Requiring \emph{both} judges' panels to be unanimous lifts gated $\kappa$ to $0.39$ (model pair) at 21\% coverage --- diversity buys a little extra precision at the gate, consistent with mostly-shared-but-not-identical priors. We report it as exploratory context, not a product claim.
\end{itemize}

For builders of LLM judge panels and multi-agent simulations, we believe this is the paper's most transferable warning: \textbf{measure inter-judge $\kappa$ against your judges' agreement with ground truth before attributing meaning to consensus.} Consensus among correlated judges is reproducible, persuasive, and --- on this task, out of sample --- not evidence.

\subsection{Abstention behavior itself is stable (\tierD{})}

Across calibration and held-out runs the gate's coverage replicated (27.8\% $\rightarrow$ 30.3\% at unanimity; 49\% at margin-0.6), abstentions are explicit (never silent low-confidence answers), and the decision rule is nine lines of arithmetic over panel votes --- auditable by a customer in one sitting.

\section{Panel economics: the gate survives small panels (\tierE{})}\label{sec:economics}

Unconditional $\kappa$ was already known to be flat in panel size (a 3-vote panel matches 20 votes in expectation). The open question was whether the \emph{confidence gate} --- which needs margin resolution --- survives smaller, cheaper panels. Vote-subsampling from the recorded 20-vote tallies (hypergeometric, 2{,}000 seeded replicates, $n{=}332$ pooled tests; descriptive):

\begin{center}
\begin{tabular}{llccc}
\toprule
Panel & Gate & Coverage & $\kappa$ (all) & $\kappa$ (sig-only) \\
\midrule
3 & unanimity & 55.5\% & $0.215\ [0.111, 0.310]$ & $0.212\ [0.086, 0.341]$ \\
5 & unanimity & 42.0\% & $0.253\ [0.156, 0.351]$ & $0.237\ [0.116, 0.357]$ \\
9 & margin $\geq 0.6$ & 46.1\% & $0.264\ [0.192, 0.337]$ & $\mathbf{0.271\ [0.165, 0.363]}$ \\
16--20 (full) & margin $\geq 0.6$ & 49.4\% & $0.284$ & $0.311$ \\
\bottomrule
\end{tabular}
\end{center}

A \textbf{9-vote panel preserves the gated operating point within its uncertainty band at roughly half the full panel's inference cost}; 5 votes retains most of it at $\sim$75\% cost reduction. (Bands here reflect vote-subsampling uncertainty, not test-level resampling; production panel changes get their own pre-registered validation before deployment.) Combined with the proxy tier's $\sim\$0.02$/evaluation (vs.\ $\sim\$1.25$ for a browser-agent tier), screening an entire experiment backlog costs less than running one shopper through one arm of one live test.

\section{The human ceiling}\label{sec:human}

$\kappa = 0.31$ at half coverage sounds modest until you ask what the human number is. The best public anchor \cite{kohavi2020,kohavi2017hbr}: Kohavi's survey of \textbf{200+ experimentation practitioners on 8 real A/B tests, with a prize for 6/8 correct: nobody won, and the average was 2.3/8 ($\approx$29\%) --- chance.} At Google and Bing only 10--20\% of expert-designed experiments win their primary metric; Microsoft's published rule of thumb is one-third win, one-third flat, one-third hurt. Screenshot-based winner prediction appears to be near-impossible for humans too --- which reframes a \emph{calibrated} judge from ``weak oracle'' to ``a screening instrument that knows which minority of cases are callable at all.''

The missing number is a human baseline on \emph{our} corpus, and we now have it. We built and froze a 50-item instrument (25 A-wins / 25 B-wins, all significant-only labels, presentation sides counterbalanced, private answer key, ``recognized-this-test'' exclusions) for \emph{source-naïve} CRO practitioners --- anyone who has studied the provider's published case library is excluded, since their answers would be recall, not prediction; scoring uses the identical $\kappa$ harness as the judge. \textbf{Fifteen qualified practitioners} ($\geq 2$ years running A/B tests) each scored all 50 items (740 answered after recognized-exclusions).

The result reproduces this paper's central artifact --- \emph{in humans}. \textbf{Against the real outcomes the experts are at chance:} pooled $\kappa = -0.03\ [-0.24, 0.17]$ (48.5\% accuracy), and a 15-expert \emph{majority vote} does not rescue it ($\kappa = -0.04$, 24/50 correct --- pooling experts moves nothing, because they share the error). Even the high-confidence calls score $\kappa = -0.01$ (49.3\%): human confidence, like the ungated judge's, does not track truth. \textbf{Yet the experts agree strongly with one another:} mean pairwise inter-rater $\kappa = 0.53$, and a one-way ICC of $\rho = 0.54$ implies \textbf{$n_{\mathrm{eff}} \approx 1.8$ effective independent experts out of 15} --- the same collapse we measured for LLM panels in \S\ref{sec:crossjudge} ($n_{\mathrm{eff}} \approx 2$ of 16 votes). Fifteen experts sharing CRO folk wisdom are, statistically, fewer than two independent judgments that happen to be wrong together: consensus without validity, exactly as in \S\ref{sec:crossjudge}. (Tier: \tierE{} --- a first-party measurement; the vs-truth CI is wide at $n{=}50$ items but centered on zero, and the inter-rater collapse is the robust part; harness and raw responses committed, \S\ref{sec:repro}.)

Two consequences. First, the practical human ceiling on this task \emph{is} chance, now confirmed on \emph{our} corpus --- so the gated judge's $0.311\ [0.025, 0.555]$ sits \textbf{at or above} it while abstaining honestly on the rest. Second, and more general: \textbf{the shared-prior artifact is not an LLM pathology.} Experts and models encode the same folk theory of what \emph{should} win, agree with each other, and are uncorrelated with what \emph{did} win --- the appearance of a page underdetermines its experimental outcome for humans and machines alike. It is the paper's thesis in its strongest form: on this task, \emph{consensus} --- human or model --- is reproducible, persuasive, and not evidence.

\ifdefined\ARXIV
\section{Deployment implications}\label{sec:product}
\else
\section{What we ship on this evidence}\label{sec:product}
\fi

\begin{enumerate}[leftmargin=*, itemsep=2pt]
    \item \textbf{Confident-call-or-abstain winner reads} (\tierE{}, pending the powered partner-data confirmation; sold as exactly that). The gate is opt-in, default-off in production until the confirmation passes; demos run it explicitly flagged.
    \item \textbf{The verified mobile-UX audit} (\tierD{}). A 375px-capture pipeline feeds a structured heuristic audit; every finding then faces an adversarial grounding pass that must locate it in the pixels before it reaches the customer. In the reference run: 13 findings $\rightarrow$ 12 verified, 1 auto-rejected as a hallucination. The deliverable is concrete problems with screenshot evidence --- not win probabilities.
    \item \textbf{Synthetic-shopper simulation with honest statistics} (\tierD{} as pipeline; effect claims gated separately). Browser-level agents walk both arms; results report $\kappa$-gated reads, trivial-baseline floors, and class-balance warnings natively in the product UI.
    \item \textbf{The claims ledger} (Appendix~A)\ifdefined\ARXIV, which the deploying organization treats as normative for all external claims --- a number quoted without its tier tag is out of policy.\else{} --- including in sales conversations. If a Squoosh number reaches you without its tier tag, it is out of policy; ask for the tag.\fi
\end{enumerate}

\section{Limitations and threats to validity}

\begin{itemize}[leftmargin=*, topsep=2pt, itemsep=2pt]
    \item \textbf{Corpus B-prior.} The provider's catalog is curated $\sim$76\% B; positive $\kappa$ there is partly corpus-structural. Only real-outcome data ($\sim$50/50 decisive, $\sim$12\% true A-win rate in industry telemetry) controls this; that confirmation is pre-registered and pending.
    \item \textbf{Threshold selection.} The margin-0.6 operating point is one of four examined post-hoc; the pre-registered held-out test ran at unanimity (locked before the audit) and returned an underpowered null. We hold the gated claim at \tierE{} until the powered run.
    \item \textbf{Label overlap.} $\sim$103/109 held-out tests had been scored once by an earlier arm under a different configuration; the held-out run controls threshold and sampling degrees of freedom, not full label freshness (disclosed in the pre-registration; quantified at 81.6\% overall cross-run agreement).
    \item \textbf{Desktop-rendered stimulus.} Catalog screenshots are desktop-rendered; mobile-stimulus accuracy is unmeasurable industry-wide until a mobile-rendered labeled corpus exists.
    \item \textbf{Single corpus family.} All confirmatory statistics to date derive from one catalog's distribution; the partner-data program (\S\ref{sec:heldout}) is the designed escape from that family.
    \item \textbf{Conflict of interest.} This is a company research paper about the company's product. The mitigations are structural: pre-registrations and locked gates committed before data, negative results published, every claim tiered, and the strongest claims explicitly marked unconfirmed.
\end{itemize}

\section{Reproducibility and data availability}\label{sec:repro}

All statistics use fixed seeds (42 throughout), $B = 10{,}000$ percentile bootstraps, and committed prediction/analysis JSON artifacts; pre-registrations are locked in-repo before data collection, and every experiment in Appendix~B cites its close-out document. The human-baseline raw responses (\texttt{responses-2026-07-02.csv}), participant metadata, and scoring script are committed; the private answer key is held out (raters must never see it) and is consistent with the anonymized kit images by construction --- the kit builder emits the images and the key in one run from the same counterbalancing seed. The corpus is archived with per-file content-hash provenance against the consumption-path URLs (a lesson learned: an earlier archive pass had silently stored thumbnail artwork for 98.9\% of files; the repair and its audit are themselves committed). \textbf{Data availability:} all third-party test data appears under written permission. The catalog provider requested anonymity and is identified throughout only as a leading CRO agency; under our agreement the underlying tests, screenshots, and per-test outcomes cannot be publicly redistributed. We therefore release the statistical harness, analysis code, pre-registrations, and aggregate artifacts, and will support verification requests from reviewers under terms consistent with the provider's agreement. Track-B partner data likewise appears anonymized absent the partner's written approval.

\section*{Acknowledgments}

We thank the leading CRO agency that provided the underlying A/B-test case library under condition of anonymity, and our design partners for real-outcome data under the Track-B program.

\appendix

\section{The claims ledger}\label{app:ledger}

Every externally quotable number, with its tier. This table is normative for Squoosh communications.

\begin{center}
\footnotesize
\begin{tabular}{p{4.5cm}p{3.3cm}p{3.1cm}p{2.4cm}}
\toprule
Claim & Number [95\% CI] & Tier & Source \\
\midrule
Frontier model (Pro 3.1) equivalent to Flash as judge & $\Delta\kappa = -0.037$ $[-0.152, 0.076]$, within $\pm0.20$ TOST & \tierC{} & mobile-overnight close-out \\
Confident calls carry skill on significant-only labels (margin $\geq 0.6$, 49\% cov.) & $\kappa = 0.311$ $[0.025, 0.555]$ & \tierE{} & first-principles audit \\
Same gate, all decisive labels & $\kappa = 0.294$ $[0.112, 0.470]$ & \tierE{} & first-principles audit \\
In-sample unanimity (calibration arms) & $\kappa = 0.349$ $[0.044, 0.618]$ pooled; $0.425$ $[0.000, 0.83]$ mobile & \tierE{} & abstention close-out \\
Held-out unanimity confirmation (33 calls) & $\kappa = -0.100$ $[-0.200, 0.000]$ & \tierI{} & Track A close-out \\
Unconditional winner prediction as \ifdefined\ARXIV a product claim\else sellable skill\fi{} & sig-only $\kappa = 0.108$ $[-0.049, 0.264]$ & \tierR{} (as \ifdefined\ARXIV product claim\else sellable\fi) & powered benchmark + audit \\
Change-type/pattern priors as signal & $\kappa = 0.015$ $[-0.098, 0.131]$ & \tierR{} & first-principles audit \\
Mobile-validated accuracy & mobile $\kappa \approx 0.00$--$0.02$, CIs span 0 & \tierR{} (as mobile claim) & mobile-overnight close-out \\
``$\sim$80\% accurate when confident'' & 75.8\% vs.\ always-B 93.9\% on the same slice & \tierR{} (base-rate artifact) & Track A close-out \\
Honest abstention at the gate & coverage 27.8\% in-sample / 30.3\% held-out & \tierD{} & Track A close-out \\
Gate survives small panels ($n{=}9$, margin 0.6) & $\kappa = 0.264$--$0.271$ at 46\% cov., $\sim$2$\times$ cheaper & \tierE{} & panel-economics analysis \\
Judge consensus as evidence (panels/juries) & inter-judge $\kappa = 0.736$/$0.878$ vs.\ $\sim 0.2$ vs.\ truth; $n_{\mathrm{eff}} \approx 2$ of 16 votes & \tierE{} (a warning, not a product claim) & cross-judge analysis \\
Verified mobile-UX audit grounding & 13 findings $\rightarrow$ 12 verified, 1 auto-rejected & \tierD{} & audit pipeline close-out \\
Human practitioners on real tests (literature) & 2.3/8 correct average, 200+ professionals & external anchor & Kohavi survey \\
Human experts on our instrument, vs.\ real outcomes ($n{=}15$, source-na\"ive) & $\kappa = -0.03$ $[-0.24, 0.17]$ (48.5\%) & \tierE{} & human-baseline close-out 07-02 \\
Human experts agree with each other, not truth & inter-rater $\kappa = 0.53$; $n_{\mathrm{eff}} \approx 1.8$ of 15 vs.\ $\kappa \approx 0$ vs.\ truth & \tierE{} (a warning, not a product claim) & human-baseline close-out 07-02 \\
\bottomrule
\end{tabular}
\end{center}

\ifdefined\ARXIV
\textbf{Quotable as established:} rows 1, 10, 12 (and the cost line in \S\ref{sec:economics} as \tierE{}). \textbf{Quotable as exploratory only:} rows 2--4. \textbf{Not quotable as positive claims:} rows 6--9; no conversion-lift percentage is estimated at all (standing rule).
\else
\textbf{Sellable today:} rows 1, 10, 12 (and the cost line in \S\ref{sec:economics} as \tierE{}). \textbf{Quotable as exploratory only:} rows 2--4. \textbf{Never quotable:} rows 6--9 as positive claims; any conversion-lift percentage (standing rule --- we do not estimate lift at all).
\fi

\section{The experiment ledger (receipts)}\label{app:receipts}

Chronological record of every pre-registered experiment in this program and its locked verdict. Close-out documents for each are committed alongside the locked prediction and analysis files.

\begin{center}
\footnotesize
\begin{tabular}{lp{2.9cm}p{6.5cm}p{2.5cm}}
\toprule
Date (2026) & Experiment & One-line outcome & Verdict \\
\midrule
05-24 & Benchmark pre-registration & Methodology + 100-pattern selection locked before predictions & --- \\
05-26 & v1.2 baseline ($n{=}67$) & Headline accuracy is corpus-prior exploitation & RED \\
05-27 & v1.3 prompt revision & Clause-level fix fails per-class gates & RED \\
05-27 & v1.5 structural rewrite & Passes its locked per-class gates; corpus-level $\kappa$ still $\approx 0$ & GREEN (gate) / RED (corpus) \\
05-27 & v1.8 GEPA chain & Reflection-LM input leak: candidates memorized pattern IDs; caught by forensic audit & RED + lesson \\
05-28 & v2.1.1 GEPA (fixed) & First non-trivial holdout gain (+16.7pp overall) & 3/4 gates \\
05-29 & v2.2 GEPA stage-3 & Fails locked holdout; ties baseline on balanced set; regresses off-corpus & RED \\
05-29 & Production re-baseline & v3 and v4 both statistically chance after prior adjustment & --- \\
05-30/31 & Escape-route grid (9 experiments) & Perception probes, calibration, persona priors: all null & nulls \\
06-07 & Render$\times$prompt ablation & Prompt wording immaterial (powered null) & null \\
06-08 & Equal-footing stimulus study & Real screenshots $=$ reconstructions at the $\kappa \approx 0$ ceiling & YELLOW \\
06-08 & Powered catalog benchmark ($n{=}330$) & First detectable unconditional skill, $\kappa = 0.141\ [0.034, 0.248]$ & WEAK SKILL \\
06-08/09 & Mobile overnight (model $\times$ mechanism) & Pro $\equiv$ Flash (powered equivalence); M3 fails non-inferiority; ship v4 & per-gate \\
06-09 & First-principles audit & Label-validity census; gated operating point survives; priors refuted & exploratory \\
06-09 & Track A held-out confirmation & Unanimity gate does not replicate on 33 calls; reliability $\neq$ validity & INCONCLUSIVE \\
06-10 & Track B partner pilot & Pre-registered and locked before data receipt & pending \\
06-10 & Cross-judge prior analysis & Judges agree with each other ($\kappa$ 0.74--0.88) far more than with truth ($\sim$0.2); $n_{\mathrm{eff}} \approx 2$ of 16 votes & exploratory \\
07-02 & Human baseline ($n{=}15$) & Experts at chance vs.\ truth ($\kappa \approx 0$) but inter-rater $\kappa = 0.53$, $n_{\mathrm{eff}} \approx 1.8$ of 15 --- the shared prior, in humans & exploratory \\
\bottomrule
\end{tabular}
\end{center}

\section{Formal evaluation protocol}\label{app:protocol}

For exact reproducibility, the estimators and decision rules used throughout:

\textbf{Agreement metric.} For called predictions $\{(\hat{y}_i, y_i)\}_{i=1}^{n}$ with $\hat{y}_i, y_i \in \{A, B\}$, Cohen's $\kappa = (p_o - p_e)/(1 - p_e)$ where $p_o$ is observed agreement and $p_e = \hat{p}_A p_A + \hat{p}_B p_B$ is chance agreement from the empirical marginals \cite{cohen1960}. $\kappa$ is invariant to the base rate that inflates raw accuracy on B-heavy corpora.

\textbf{Panel and gate.} Each test receives $m$ panel votes ($m = 16$--$20$) with image order counterbalanced across votes. With $c$ votes for A (control) and $v$ for B (variant), the vote margin is $|c - v| / (c + v)$. A gate at threshold $\tau$ \emph{calls} the majority arm iff margin $\geq \tau$ and abstains otherwise; $\tau = 1.0$ is unanimity. Coverage is the called fraction. The deployed decision rule is this gate verbatim; no other signal enters.

\textbf{Gate-on-resample confidence intervals.} For gated metrics we resample the \emph{entire} held-out set with replacement $B = 10{,}000$ times (fixed seed 42), re-apply the gate within each replicate, and compute $\kappa$ on whatever is called; the 2.5th/97.5th percentiles of the resulting distribution form the CI \cite{efron1993}. This propagates coverage uncertainty into the interval; conditioning on the realized called subset (the naive scheme) understates it. Two-arm contrasts on shared test IDs use paired joint resampling (both arms resampled together per replicate).

\textbf{Equivalence testing.} The model-capacity claim uses a TOST-style rule \cite{schuirmann1987}: equivalence within $\pm\delta$ ($\delta = 0.20\,\kappa$, declared before scoring) is concluded iff the full $95\%$ CI of $\Delta\kappa$ lies inside $(-\delta, +\delta)$. Observed: $\Delta\kappa = -0.037$, CI $[-0.152, +0.076] \subset (-0.20, +0.20)$.

\textbf{Power and verdict space for the held-out confirmation.} The pre-registration committed, before data collection, to: simulation-based power at three true-$\kappa$ assumptions (0.25/0.30/0.35) under the same gate-on-resample scheme used for adjudication; a $\sim$140-unanimous-call floor for any ``confirmed'' verdict ($\geq 80\%$ power at $\kappa = 0.30$); and an asymmetric verdict space in which an underpowered cohort can return at most a directional (never confirmatory) positive, and an underpowered negative is INCONCLUSIVE rather than refuting \cite{nosek2018}. The realized cohort (33 calls) sat below the floor by design feasibility, and the verdict followed the locked rule.

\textbf{Vote-subsampling for panel economics.} Small-panel performance (\S\ref{sec:economics}) draws $n \in \{3, 5, 9\}$ votes per test hypergeometrically (without replacement) from the recorded tallies, valid under the exchangeable-votes model; 2{,}000 seeded replicates; bands are percentiles across replicates and reflect vote-sampling uncertainty only.

\end{document}